\documentclass[a4paper]{iopart}

\usepackage[dvips]{graphicx}
\usepackage{times}
\usepackage{latexsym}
\usepackage{amssymb}
\usepackage{color}

\begin{document}

\title{How to tell a gravastar from a black hole}

\author{Cecilia B M H Chirenti$^{1,2}$ and Luciano Rezzolla$^{2,3}$}

\address{$^1$ Instituto de F\'\i sica, Universidade de S\~ao Paulo, 
S\~ao Paulo, Brazil}

\address{$^2$ Max-Planck-Institut f\"ur Gravitationsphysik, Albert Einstein 
Institut, 
14476 Golm, Germany}

\address{$^3$ Department of Physics, Louisiana State University, Baton
  Rouge, LA 70803 USA}

\begin{abstract}
  Gravastars have been recently proposed as potential alternatives to
  explain the astrophysical phenomenology traditionally associated to
  black holes, raising the question of whether the two objects can be
  distinguished at all. Leaving aside the debate about the processes
  that would lead to the formation of a gravastar and the astronomical
  evidence in their support, we here address two basic questions: Is a
  gravastar stable against generic perturbations? If stable, can an
  observer distinguish it from a black hole of the same mass?  To
  answer these questions we construct a general class of gravastars
  and determine the conditions they must satisfy in order to exist as
  equilibrium solutions of the Einstein equations. For such models
  we perform a systematic stability analysis against
  axial-perturbations, computing the real and imaginary parts of the
  eigenfrequencies. Overall, we find that gravastars are stable to
  axial perturbations, but also that their quasi-normal modes differ
  from those of a black hole of the same mass and thus can be used to
  discern, beyond dispute, a gravastar from a black hole.
\end{abstract}
\maketitle

\section{Introduction}

The gravastar model was proposed by Mazur and Mottola~\cite{Mazur} and
has attracted attention as a possible alternative to black
holes. Within this model, in fact, a massive star in its late stages
could end its life as a gravastar, namely a very compact object whose
radius would be very close to the Schwarzschild radius (indeed
arbitrarily close to it) without having an event horizon or a central
singularity. For this to happen, a phase transition is expected to
take place at or near the location where the event horizon would have
been formed otherwise~\cite{Chapline}. The interior of what would have
been the black hole is replaced by a suitably chosen portion of
de-Sitter spacetime with an equation of state (EOS) $p=-\rho$,
surrounded by a thin shell of ultra-stiff matter with EOS $p=+\rho$,
and which is then suitably matched to a Schwarzschild vacuum. In
practice, the Mazur-Mottola gravastar model, hereafter the MM
model~\cite{Mazur}, is a spherically symmetric and static five-layer
solution of the Einstein equations, including two infinitesimally thin
shells needed by the junction conditions of the metric~\cite{Israel}.

For as much as it is ingenious, the gravastar model also challenges a
building block in modern astronomy, namely the existence of
astrophysical black holes. In spite of the many observational
evidences in favour of black holes, it may indeed be fundamentally
impossible to give an irrefutable observational proof for the
existence of a black-hole horizon if only electromagnetic radiation is
received~\cite{Abramowicz} and thus tell apart a black hole from a
gravastar, if the latter existed. Such a challenge has obviously
attracted a lot of attention and several related models have been
recently proposed in the attempt of providing answers to two basic
questions: Once produced, is a gravastar stable to generic
perturbations?  If stable, can an external obverver distinguish a
gravastar from a black hole of the same mass?

In ref.~\cite{Mazur} it was argued that the MM solution is
thermodynamically stable, but other stability analysis are not so easy
to perform because of the model's structure.  Visser and
Wiltshire~\cite{Visser} have analyzed the radial stability of a
simplified model with three layers and the stability was shown to hold
for a number of configurations. This stability was generalized by
Carter~\cite{Carter} for gravastar models with different
exteriors. Other possibilities for the interior solution have also
been considered: Bili\'c \etal \cite{Bilic} replaced the de-Sitter
interior by a Born-Infeld phantom, while Lobo replaced the interior
solution by one that is governed by the dark-energy EOS~\cite{Lobo1},
and Lobo and Arellano matched interior nonlinear electrodynamic
geometries to the Schwarzschild exterior~\cite{Lobo2}.

To remove in part the complications produced by the infinitesimal
shells in the MM model, Cattoen \etal~\cite{Cattoen} have found that
fluid gravastars can be built if the fluid is confined to a given
layer and has there anisotropic pressures. The latter essentially
replace the surface tension which was introduced in the original model
MM model by the matching of the metric in the infinitesimally thin
shells. Although anisotropic and with rather arbitrary equations of
state, these pressures have the appealing property of being continuous
and thus of allowing one to build equilibrium models without the
presence of infinitesimally thin shells and thus look more seriously
into the issue of stability. Indeed, as a step to go beyond the
construction of equilibrium models, DeBenedictis \etal
\cite{DeBenedictis} made a first attempt to investigate the stability
of gravastars through a qualitative analysis of axial perturbations.

In order to provide more definitive answers to the questions mentioned
above, we have constructed a general class of fluid gravastars with
finite shell thickness and variable compactness. While this class is
similar and has been inspired by the one proposed
in~\cite{Cattoen,DeBenedictis}, it also differs in two important
aspects. Firstly, while we also consider a fluid gravastar with
anisotropic pressures, we try and reproduce the most salient features
of the original MM model by creating an internal and an external region
which reproduce a de-Sitter and a Schwarzschild spacetime at finite
radii and not only asymptotically. Secondly, for these models we
determine precise bounds for the properties of the metric functions
and the compactness of the gravastar that yield equilibrium solutions,
thus restricting considerably the class of possible solutions.

Using these models we have carried out a systematic investigation of
the stability analysis of gravastars against
axial-perturbations~\cite{Chandra1, Chandra2}, thus extending the
results discussed in~\cite{Dymnikova} to our gravastar model and
computing explicit eigenfrequencies. In this way we were able to
conclude that: a gravastar \textit{is stable} to axial perturbations
and indeed \textit{it is possible} to distinguish it from a black hole
if gravitational radiation is produced. More specifically, we have
found that for all the models considered the imaginary part of the
eigenfrequencies is always negative, thus indicating stability against
these perturbations. Furthermore, while it is always possible to build
a gravastar with given compactness and thickness such that it will
have the same oscillation frequency as that of a black hole with the
same mass, the corresponding decaying time will be different. Our
results thus provide a way of distinguishing observationally and
beyond dispute a gravastar from a black hole.

The paper is organized as follows: in section 2 we review the main
features of the original MM model and obtain numerical solutions which
enable us to impose bounds for the parameters of the MM solution. In
section 3, instead, we present our model for a fluid gravastar with
anisotropic pressures and discuss the bounds derived from general
conditions imposed by the EOS and the properties of the metric
functions. In Section 4 we outline the perturbation equations and the
numerical methods employed in the study of axial perturbations of our
model.  In section 5 we discuss the results obtained for the QNMs and
in section 6 we present our final conclusions.

\section{The MM gravastar model}

We first present a quick review on the main features of the original
MM model~\cite{Mazur} for they will be used in further developments in
this paper and highlight that precise bounds exist for the existence
of solutions even for the simplest gravastar model.

In general, we consider a static and spherically symmetric line
element
\begin{equation}
  \rmd s^2 = -f(r)\rmd t^2+\frac{\rmd r^2}{h(r)}+r^2\rmd\Omega^2\,,
\end{equation}
and the Einstein equations must be solved for a perfect fluid at rest,
such that there are three different regions with the three different
equations of state
\begin{eqnarray}
\begin{array}{rlcl}
\textrm{I.} & \textrm{Interior:}\; & 0 \le r \le r_1\,,\; & \rho =-p\,,\\ \\
\textrm{II.} & \textrm{Shell:}\; & r_1 \le r \le r_2\,,\; & \rho =+p\,,\\ \\
\textrm{III.} & \textrm{Exterior:}\; & r_2 \le r\,,\; & \rho = p = 0\,.
  \end{array}
\end{eqnarray}

In region I, $\rho$ is a constant given by $\rho_v = 3H_0^2/8\pi$, and
the metric is that of a de-Sitter spacetime, so
\begin{equation}
f(r) = Ch(r) = C(1-H_0^2r^2)\,,\qquad 0 \le r \le r_1\,.
\end{equation}
where $C$ is an integration constant, whose value will be determined
later [\textit{cf.} eq.~(\ref{int_const})].

In region II, a new dimensionless variable $w$ is defined as $w \equiv
8\pi r^2p$, in order to obtain the following set of equations
\begin{eqnarray}
\label{rescaled}
  \frac{\rmd r}{r} = \frac{\rmd h}{1-w-h}\,,\\
  \frac{\rmd h}{h} = -\left(\frac{1-w-h}{1+w-3h}\right)\frac{\rmd w}{w}\,,\\
  \frac{wf}{r^2} = \textrm{const.}
\end{eqnarray}
An analytical solution can be obtained in the case of a very thin
shell, \textit{i.e.}, for $r_1\to r_2$. It is easy to see that
(\ref{rescaled}) is satisfied if we take $h = 1-{\bar m}/r$ and $\rmd
{\bar m}(r) = 2\rmd m(r) = 8\pi\rho r^2\rmd r^2 = w\rmd r$. Therefore,
in the thin-shell limit, (\ref{rescaled}) can be integrated to yield
\begin{equation}
h \equiv 1 - \frac{{\bar m}}{r} \simeq \epsilon\frac{(1+w)^2}{w} \ll 1\,,
\end{equation}
where $\epsilon$ is an integration constant. From the continuity of
the metric coefficients $f$ and $h$ at $r_1$ and $r_2$, it can be
shown that the integration constants $\epsilon$, $C$, $M$ and $H_0$
are given in terms of $r_1$, $r_2$, $w_1$ and $w_2$ by the relations
\begin{eqnarray}
\label{int_const}
  \epsilon = -\ln \frac{r_2}{r_1}\left(\ln \frac{w_2}{w_1}-\frac{1}{w_2}+
    \frac{1}{w_1}\right)^{-1}\,,\\
  C = \left(\frac{1+w_2}{1+w_1}\right)^2\,,\\
  M = \frac{r_2}{2}\left[1-\frac{\epsilon (1+w_2)^2}{w_2}\right]
    = m(r_2)\,,\\
  H_0^2 = \frac{1}{r_1^2}\left[1-\frac{\epsilon (1+w_1)^2}{w_1}\right]\,.
\end{eqnarray}

Finally, in region III and because of Birkhoff's theorem, the
spacetime is described by the Schwarzschild metric,
\begin{equation}
f(r) = h(r) = 1-\frac{2M}{r}\,,\qquad r_2 \le r\,.
\end{equation}

Clearly, region II is the most interesting one from a physical point
of view since it is where a non-trivial model for the gravastar can be
specified. In the case of a ``thick shell'', that is away from the
thin-shell limit, the gravastar solution in region II can be obtained
in the same way as for ordinary spherical stars, namely through the
numerical solution of the Tolman-Oppenheimer-Volkoff (TOV) equations
for $\rho(r) = p(r)$ and $m(r)$. We recall that in the MM model
$\rho(r)$ and $p(r)$ are discontinuous in $r_1$ and $r_2$ and that
$m(r_1)$ must, of course, be less than $r_1/2$

By specifying $r_1$, $r_ 2$, and thus the shell thickness $\delta
\equiv r_2 - r_1$, as well as the initial conditions for the pressure
$p(r_1)$ and mass $m(r_1)$, the numerical solution of the TOV
equations provides $m(r)$ and $\rho(r)$ for $r_1 < r \le r_2$. A
systematic analysis has revealed that a limit exists on the
compactness $\mu \equiv M/r_2$ of the gravastar and in particular
that, for given $r_1$, $r_2$ and $m(r_1)$ very close to the limit
$r_1/2$, there exists a value for $\rho(r_1)$ which provides the
largest $M$. In other words, it is not possible to achieve arbitrary
values for $M$. On the other hand, if we fix $r_2$ and decrease $r_1$
so as to increase the thickness of the shell, we must also decrease
$m(r_1)$, and we have verified that the maximum value obtained for $M$
also decreases. Overall, therefore, each shell thickness $\delta$ also
selects a maximum compactness of the gravastar $\mu$, above which no
solution of the TOV equations can be found. 

\begin{figure}[htp!]
  \includegraphics[angle=270,width=0.8\linewidth]{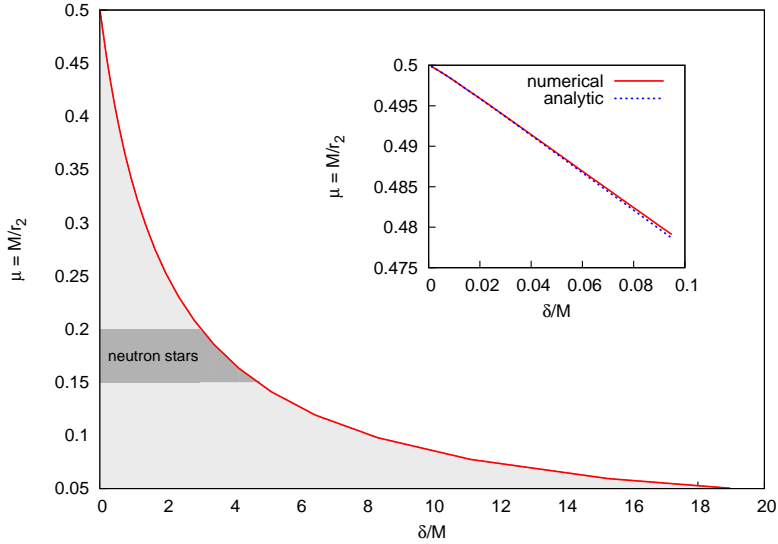}
  \caption{Limits in the compactness and thickness of a gravastar in
    the MM model. The curve shows the maximum compactness for a given
    thickness $\delta$ of the shell, that is, in the light gray area
    below the curve we have the possible solutions, while in the area
    above the curve solutions are no longer possible. The dark gray
    area highlights the typical compactnesses for neutron stars. The
    inset shows a comparison between the numerical solution of the TOV
    equations with the analytical solution in the thin-shell
    limit.}
  \label{graf_novo}
\end{figure}

All of this is shown in figure~\ref{graf_novo} which reports the space
of parameters $(\delta,\mu)$ where equilibrium solutions can be
found. The solid curve, in particular, is computed numerically and
 distinguishes the region where equilibrium models can be found
(\textit{i.e.}, the region below the curve), from the region where no
solutions can be found (\textit{i.e.}, the region above the
curve). Stated differently, for any given shell thickness $\delta$,
the solid curve marks the maximum compactness (\textit{i.e.}, the
largest value of $\mu$) for which gravastar models can be
built. Interestingly, therefore, it is not possible to build a very
compact gravastar with a very thick matter shell. Rather, gravastars
have either large compactness and thin shells, or small compactness
and thicker shells, as shown in figure~\ref{graf_novo}. This figure, and
its inset in particular, also illustrates that gravastars can be built
with arbitrarily large compactness, \textit{i.e.,} with $\mu
\rightarrow 1/2$ and thus with the outer radius $r_2$ being only
infinitesimally larger than the corresponding Schwarzschild radius. It
is exactly this property that makes gravastars hard to distinguish
from a black hole if only electromagnetic radiation is available.

\section{The gravastar model with anisotropic pressures}

\subsection{The field equations}

We now discuss our suggestion for a fluid gravastar model in which we
follow the spirit of the original MM model and consider a spacetime
consisting of three different regions with the internal and external
ones reproducing a de-Sitter and a Schwarzschild spacetime at finite
radii, $r_1$ and $r_2$, respectively. In addition, we follow the
suggestion made in~\cite{Cattoen} and use a thick shell with a
continuous profile of anisotropic pressures to avoid the introduction of
infinitesimally thin shells of matter as demanded by the
MM~\cite{Mazur}. We start therefore with a metric of the form
\begin{equation}
  \rmd s^2 = -\rme^{\nu(r)}\rmd t^2+\rme^{\lambda(r)}\rmd r^2+r^2\rmd\Omega^2
  \,,
\end{equation}
and a fluid stress-energy tensor $T^{\mu}_{\phantom{\mu}\nu} =
\textrm{diag}[-\rho, p_r, p_t, p_t]$, where $p_r$ and $p_t$ are the
radial and tangential pressures, respectively. The Einstein field
equations for this spacetime geometry and matter distribution are
\begin{equation}
  \rme^{-\lambda}\left(\frac{\lambda '}{r}-\frac{1}{r^2}\right)+\frac{1}{r^2} 
  = 8\pi \rho\,,
  \label{field1}
\end{equation} 
\begin{equation}
  \rme^{-\lambda}\left(\frac{\nu '}{r}+\frac{1}{r^2}\right)-\frac{1}{r^2} = 
  8\pi p_r\,,
  \label{field2}
\end{equation} 
\begin{equation}
  \rme^{-\lambda}\left(\frac{\nu ''}{2}-\frac{\lambda '\nu '}{4}+
  \frac{\nu '^2}{4}+\frac{\nu '-\lambda '}{2r}\right) = 8\pi p_t\,.
  \label{field3}
\end{equation} 
It is now convenient to transform the above equations into a form
where the hydrodynamical properties of the system are more evident and
that reduces to the TOV equations for systems with isotropic pressure,
\textit{i.e.},

\begin{equation}
  \rme^{-\lambda} = 1-\frac{2m(r)}{r}\,,
  \label{lambda}
\end{equation}
\begin{equation}
  \nu ' = \frac{2m(r)+8\pi r^3p_r}{r(r-2m(r))}\,,
  \label{nu'}
\end{equation}
\begin{equation}
  p_r' = -(\rho+p_r)\frac{\nu '}{2}+\frac{2(p_t-p_r)}{r}\,,
  \label{p_r'}
\end{equation}
where 
\begin{equation}
  m(r) \equiv \int_0^r 4\pi r^2\rho \rmd r\,.
\end{equation}
Combining (\ref{nu'}) and (\ref{p_r'}), we obtain the anisotropic TOV equation
\begin{equation}
p_r' = -(\rho+p_r)\frac{m(r)+4\pi r^3p_r}{r(r-2m(r))}+\frac{2(p_t-p_r)}{r}\,,
  \label{aTOV}
\end{equation}
which is reminescent of the Newtonian hydrostatic-equilibrium equation
and where the last term is obviously zero in the case of isotropic
pressures, \textit{i.e.}, $p_t=p_r$.

\subsection{Equation of state}

\begin{figure}[htp!]
  \includegraphics[angle=270,width=0.5\linewidth]{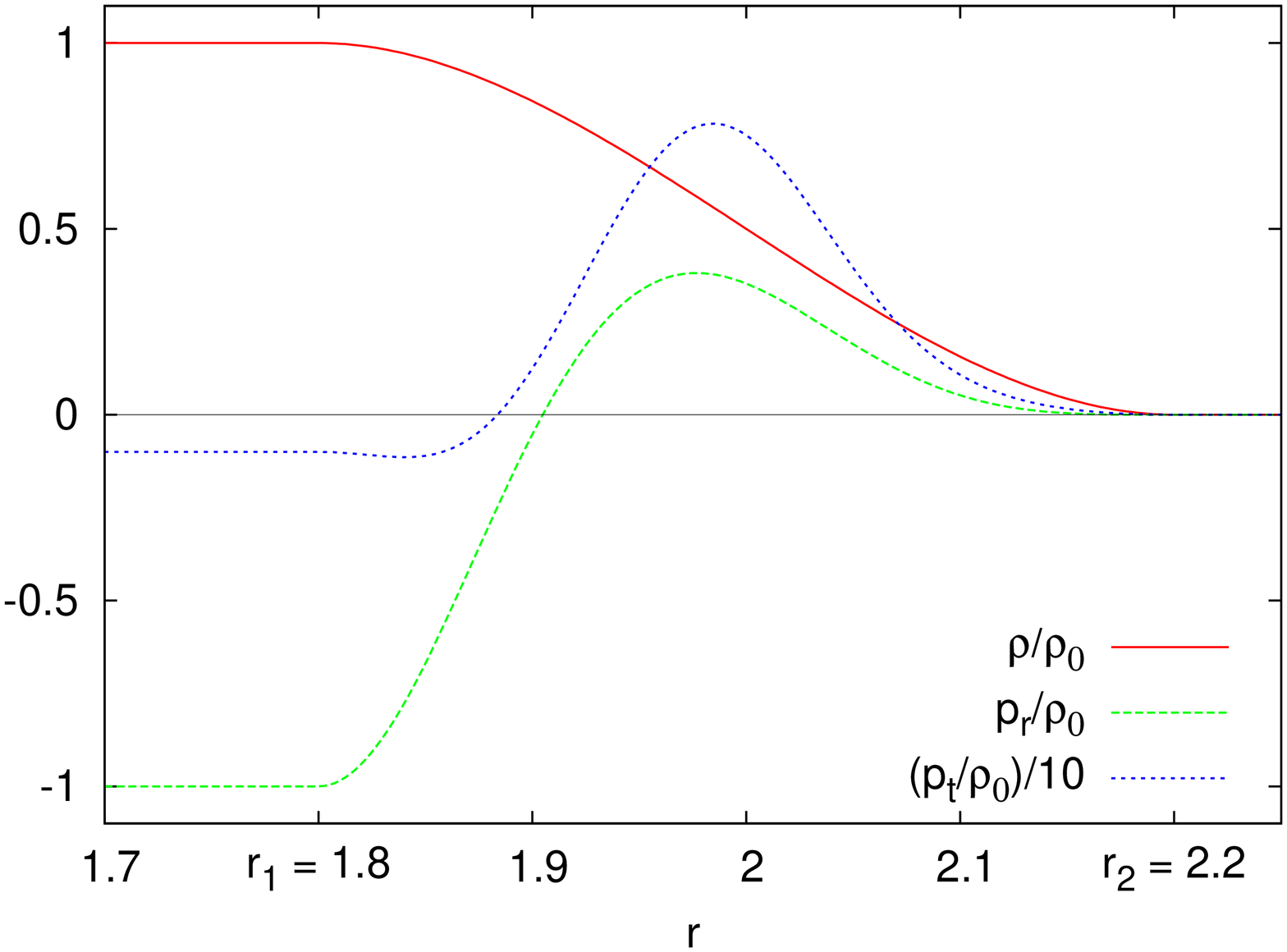}
  \hskip 0.2cm
  \includegraphics[angle=270,width=0.5\linewidth]{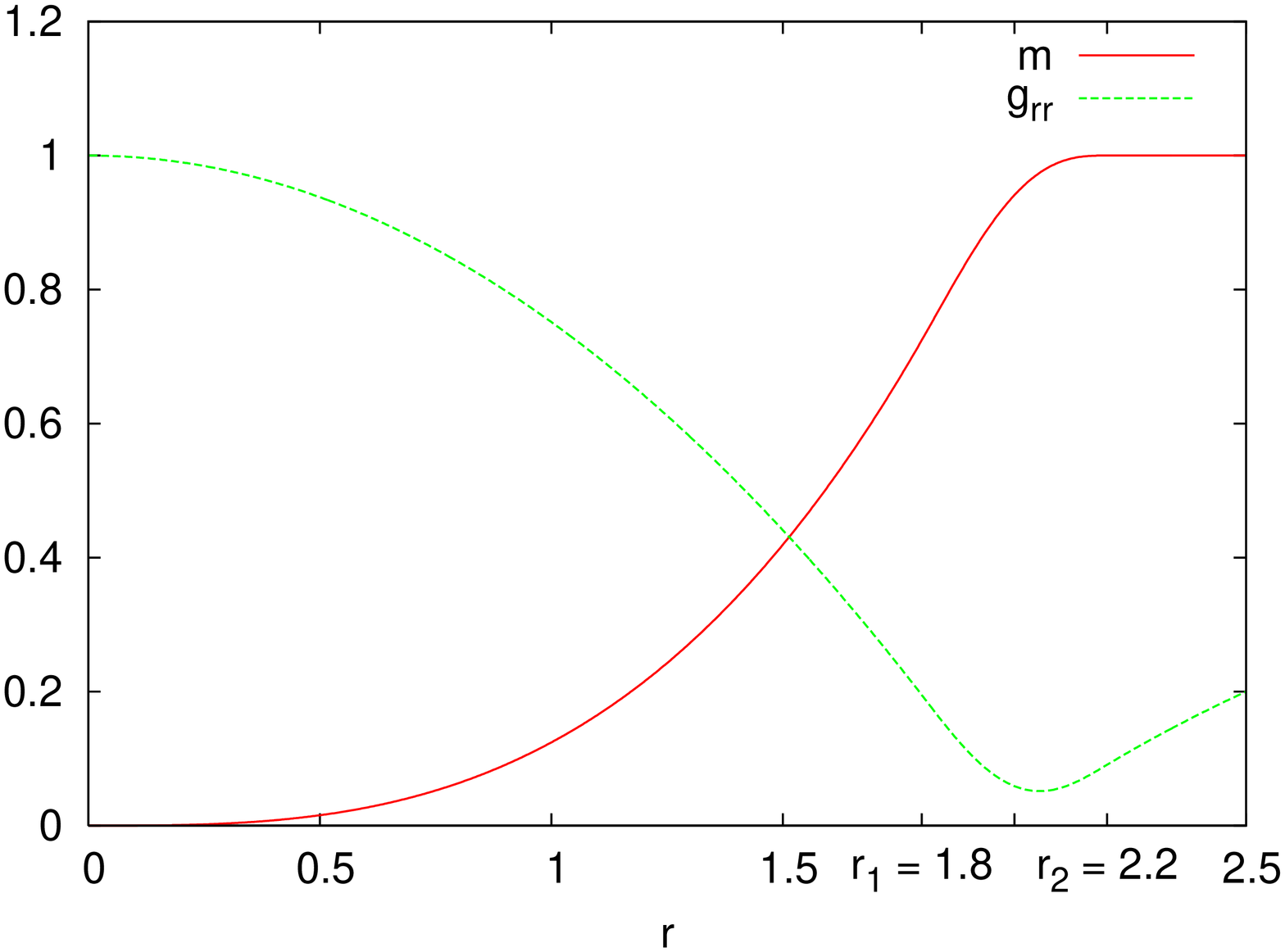}
  \caption{\textit{Left panel:} Behaviour of the functions $\rho(r)$,
    $p_r(r)$ and $p_t(r)$ for a representative gravastar model with $M
    = 1$, $r_1 = 1.8$ and $r_2 = 2.2$. The function $p_t(r)$ is scaled
    by 0.1 for better visualization but provides the dominant
    contribution in eq.~(\ref{aTOV}). \textit{Right panel:} Behaviour
    of the mass function $m(r)$ and of the metric coefficient $g_{rr}$
    for the same solution shown in the left panel.}
  \label{eos1}
\end{figure}

In order to adopt the model suggested in~\cite{Cattoen}, and still
maintain the simple structure of the MM model, we make the following
choices for our density function $\rho(r)$
\begin{eqnarray}
  \rho(0) = \rho(r_1) = \rho_0,\qquad \rho(r_2) = 0\,, \qquad
  \rho'(r_1) = \rho'(r_2) = 0\,.
  \label{cond_rho}
\end{eqnarray}
A simple way to satisfy the above conditions is to consider a cubic
dependence in $r$,
\begin{equation}
\rho(r) = \left\{ \begin{array}{lll}
    \rho_0\,, & 0 \le r \le r_1       &\qquad \textrm{region I.}\\
    ar^3+br^2+cr+d\,, & r_1 < r < r_2 &\qquad \textrm{region II.}\\
    0\,, & r_2 \le r  	              &\qquad \textrm{region III.}
  \end{array} \right.\,,
\label{rho}
\end{equation}
with the coefficients $a,b,c,d$ given by
\begin{eqnarray}
  a = \frac{2\rho_0}{(r_2-r_1)^3}\,,\\
  b = -\frac{3\rho_0(r_2+r_1)}{(r_2-r_1)^3}\,,\\
  c = \frac{6\rho_0r_1r_2}{(r_2-r_1)^3}\,,\\
  d = \frac{\rho_0(r_2^3-3r_1r_2^2)}{(r_2-r_1)^3}\,.
\end{eqnarray}
To obtain the density in the terms of the total mass $m(r_2) = M$, we write 
\begin{eqnarray}
\fl \qquad \rho_0 = M\frac{(r_2-r_1)^3}{4\pi}\left[\frac{(r_2^6-r_1^6)}{3}
    -\frac{3(r_2+r_1)(r_2^5-r_1^5)}{5}+
    \frac{3r_1r_2(r_2^4-r_1^4)}{2}+{}\right.\nonumber\\
    \left.{}\hskip 2.5cm +\frac{(r_2^3-3r_1r_2^2)(r_2^3-r_1^3)}{3}
    +\frac{r_1^3(r_2-r_1)^3}{3}\right]^{-1}\,.
\end{eqnarray}
For the radial pressure $p_r$, we follow the suggestion made in
refs.~\cite{Mbonye,DeBenedictis} and use an EOS of the type
\begin{equation}
  p_r(\rho) = \left(\frac{\rho^2}{\rho_0}\right)
  \left[\alpha-(1+\alpha)\left(\frac{\rho}{\rho_0}\right)^2\right]
  \,.
  \label{eos}
\end{equation}
Clearly, the EOS (\ref{eos}) cannot be derived from basic principles
and serves here essentially as a closure relation for the system of
the equations. Yet, such an EOS can be constrained and the parameter
$\alpha$ is determined by demanding that the maximum sound speed
${\rmd^2 p_r}/{\rmd\rho^2} = 0$ coincides with the speed of light to
rule out a superluminal behaviour. Because the sound speed
\begin{equation}
c_s \equiv \frac{\rmd p_r}{\rmd\rho} = 
	2\left(\frac{\rho}{\rho_0}\right)
	\left[\alpha - 2 (1+\alpha)
	\left(\frac{\rho}{\rho_0}\right)^2\right] \,,
\end{equation}
has a maximum for ${\bar \rho} = \rho_0\sqrt{\alpha/(2(1+\alpha))}$,
requiring $c_s({\bar \rho})=1$ yields $\alpha \simeq 2.2135$. Despite
being not particularly realistic, the EOS (\ref{eos}) has the
advantage of being simple and of possessing the needed physical limits
since $p_r \to -\rho$ for $r \to r_1$ and $p_r \to 0$ for $r \to r_2$
(\textit{cf.} right panel of figure \ref{eos1} for a typical gravastar
model).

Once the solutions for the energy density and the radial pressure are
known, the tangential pressure $p_t$ can be computed through the
anisotropic TOV equation (\ref{aTOV}) as
\begin{equation}
  p_t = p_r+\frac{r}{2}p_r'+\frac{1}{2}(p_r+\rho)
  \left[\frac{m(r)+4\pi r^3p_r}{r(1-2m(r)/r)}\right]\,.
  \label{p_t}
\end{equation}
The set of equations (\ref{cond_rho})--(\ref{p_t}) fully determines
our gravastar model which will have a finite core described by the
de-Sitter metric, a crust of matter and an exterior described by the
Schwarzschild solution. 

The behaviour of the energy density and of the pressures is shown in
the left panel of figure~\ref{eos1} for a representative gravastar
model with $M = 1$, $r_1 = 1.8\,M$ and $r_2 = 2.2\,M$. The right panel
on the same figure, instead, shows the mass function and the metric
coefficient $g_{rr}$ for the same representative model.

\subsection{Conditions on the metric functions}

While equations (\ref{cond_rho})--(\ref{p_t}) allow to construct
gravastar models, additional constrains need to be imposed to guarantee
that the metric functions $g_{tt}$ and $g_{rr}$ (and their first
derivatives) have the expected properties once our choice for the EOS
$p_r(\rho)$ and $\rho(r)$ is made. In particular, it is not difficult
to conclude that $m(r)$ and $m'(r)$ 
are continuous for the choice of $\rho(r)$ made in equation
(\ref{rho}), so that using equation
(\ref{lambda}) $g_{rr}$ and its first derivative are continuous
throughout the spacetime.

Similar considerations apply for the metric function $g_{tt}$, for
which we note that in obtaining the solution for $\nu$ by integrating
equation (\ref{nu'}) we must allow for a constant of integration
$\nu_0$ so that
\begin{equation}
  \nu = \int_0^r \frac{2m(r)+8\pi r^3 p_r}{r(r-2m(r))}\rmd r+\nu_0\,.
\end{equation}
The integration constant can then be determined through the condition
that, at the surface of the gravastar, $g_{tt}(r_2) =
-\left(1-{2M}/{r_2}\right)$, thus giving
\begin{equation}
  g_{tt} = -\rme^{\nu(r)}=
  -\left(1-\frac{2M}{r_2}\right)\rme^{\Gamma(r)-\Gamma(r_2)}\,,
  \label{g_00}
\end{equation}
where 
\begin{equation}
\Gamma(r) \equiv \int_0^r \frac{2m(r)+8\pi r^3 p_r}{r(r-2m(r))}\rmd r\,.
\label{Gamma}
\end{equation}
In this way $g_{tt}$ and $g_{tt}'$ are continuous across $r_2$ and
throughout the spacetime.

\begin{figure}[htp!]
  \includegraphics[angle=270,width=0.5\linewidth]{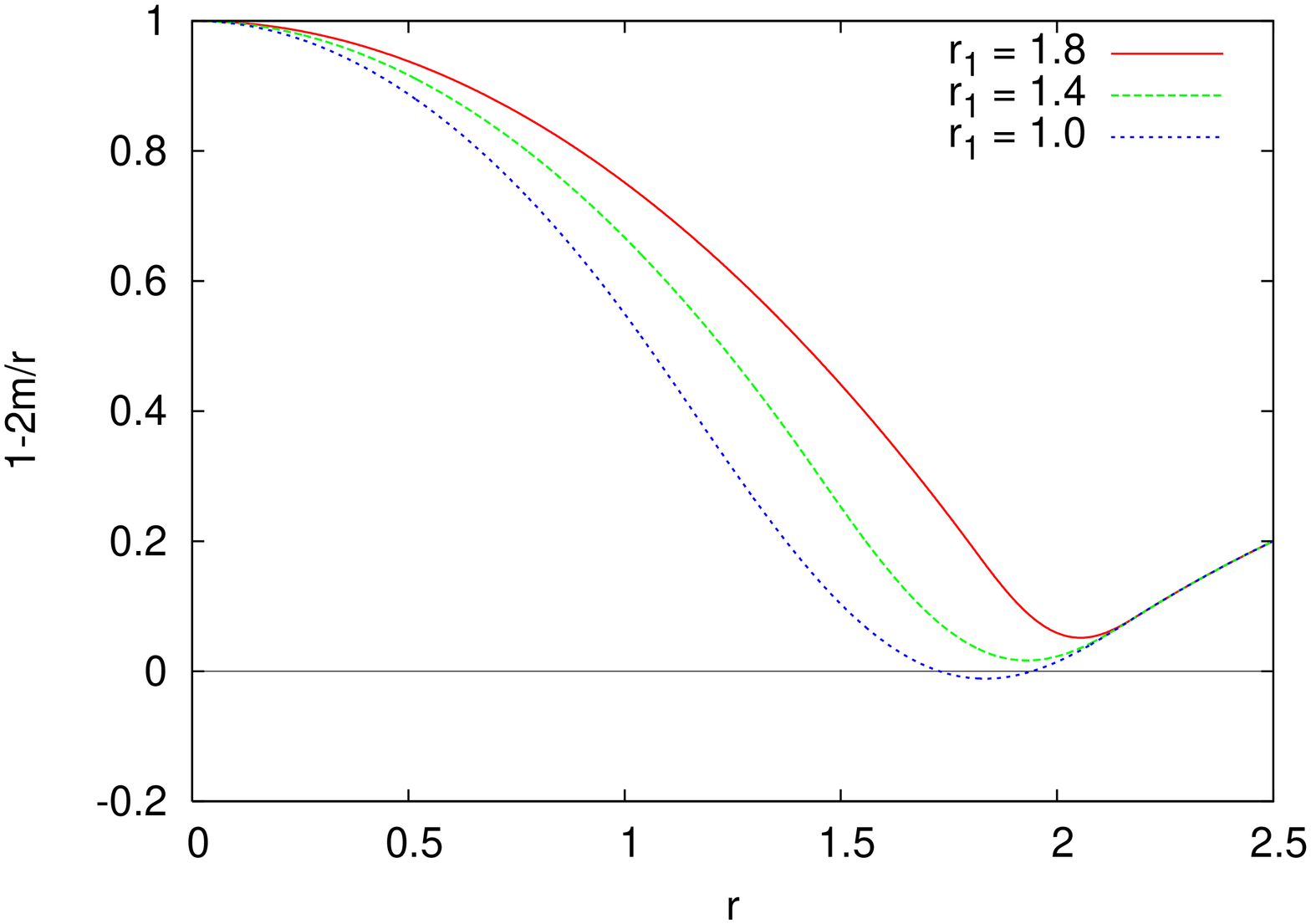}
  \hskip 0.2 cm
  \includegraphics[angle=270,width=0.5\linewidth]{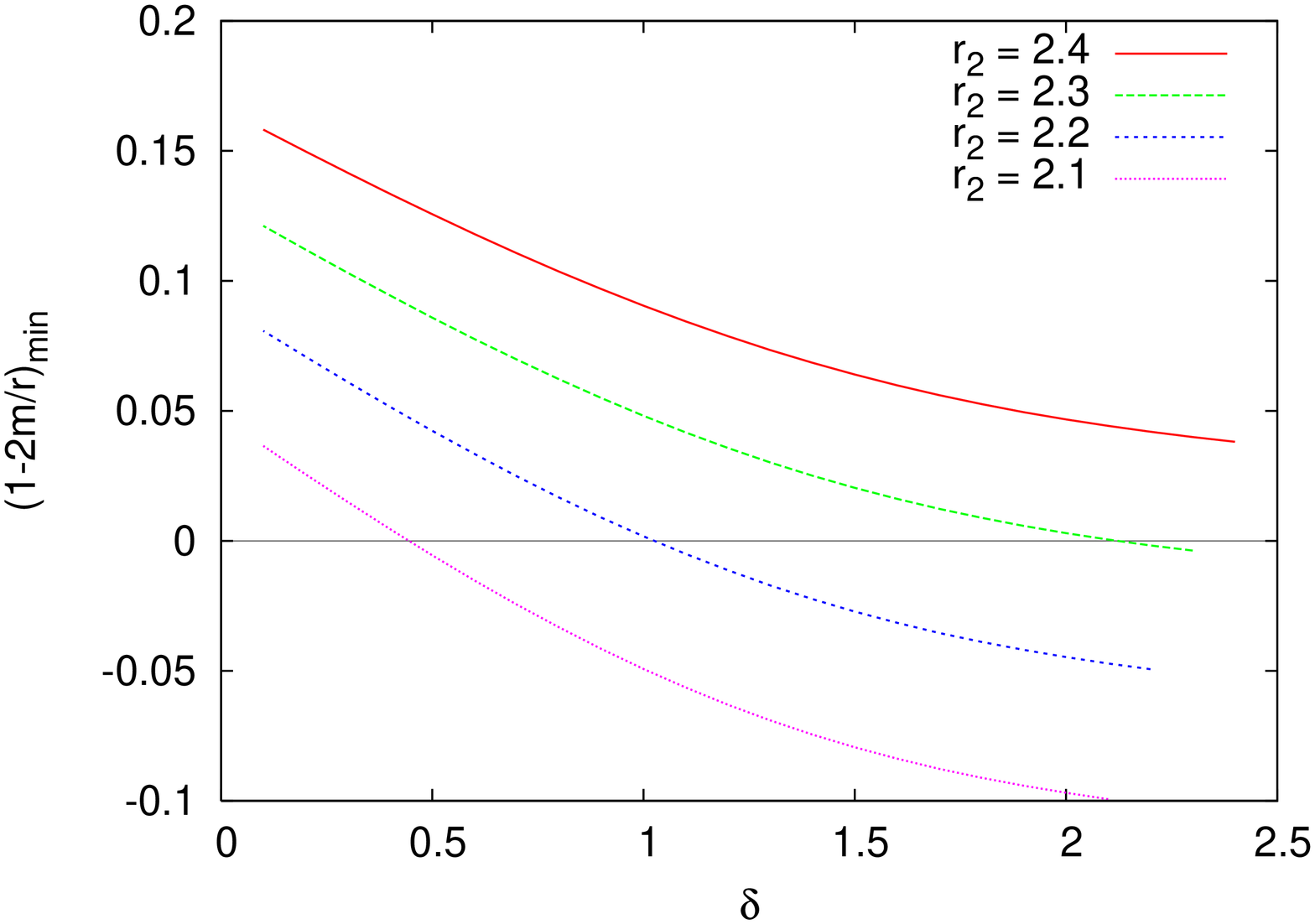}
  \caption{\textit{Left panel:} Behaviour of $1-2m(r)/r$ for $M = 1$
    and different values of $r_1$ and $r_2 = 2.2$. \textit{Right
    panel:} Dependence on $\delta$ of the minimum value of $1-2m(r)/r$
    for gravastars with $M = 1$ and different values of $r_2$. Note
    that for sufficiently large models, the minimum is always
    positive.}
  \label{f_r}
\end{figure} 

The requirement for the metric functions of being continuous with
continuous first derivatives is, in general, not sufficient to
guarantee the existence of acceptable equilibrium models. The three
free parameters $M$, $r_1$ and $r_2$, in fact, cannot be chosen
arbitrarily but in such a way that $g_{rr}$ is always positive. In the
left panel of figure~\ref{f_r} we show a typical example of the
behaviour of the function $1-2m(r)/r$, where different lines refer to
different values of the inner radius $r_1$ while the outer one $r_2$
is held fixed. Clearly, an incorrect choice of $r_1$ and $r_2$ leads
to negative values for $g_{rr}$ and thus to the unphysical appearance
of horizons. The right panel of figure~\ref{f_r} shows how the minimum
value of the function $1-2m(r)/r$ depends on the gravastar's thickness
$\delta$ and that gravastars with large thicknesses can be built if
the outer radius $r_2$ is chosen to be sufficiently large, that is, if
the compactness $\mu$ is sufficiently small.

As a more systematic characterization of this problem, we have
constructed a large number of gravastar models in which we have varied
both the compactness $\mu$ and the thickness $\delta$. In this way we
can extend the diagram presented in figure~\ref{graf_novo} for the
existence of MM models to our thick-shell gravastar model and define
the region in the $(\delta, \mu)$ parameter space where equilibrium
solutions can be found. In particular, by definining the parameter
$\epsilon$ as the dimensionless distance of the gravastar's surface
from a Schwarzschild horizon, \textit{i.e.,}~$\epsilon \equiv r_2/M -
2$, we find that for $\epsilon > \epsilon_c \approx 0.3085$, the
function $1-2m(r)/r$ is always positive, so that the thickness
$\delta$ can be as large as $r_2$ (\textit{i.e.}, $r_1$ can also be
taken to be zero). On the other hand, for $\epsilon \leq \epsilon_c$
there exists a critical thickness $\delta_c=\delta_c(r_2,M)$ above
which equilibrium models cannot be found because of the appearance of
horizons. For $\epsilon=\epsilon_c$, we have found that
$\delta_c=2.30685\,M$.

\begin{figure}[htp!]
  \includegraphics[angle=270,width=0.8\linewidth]{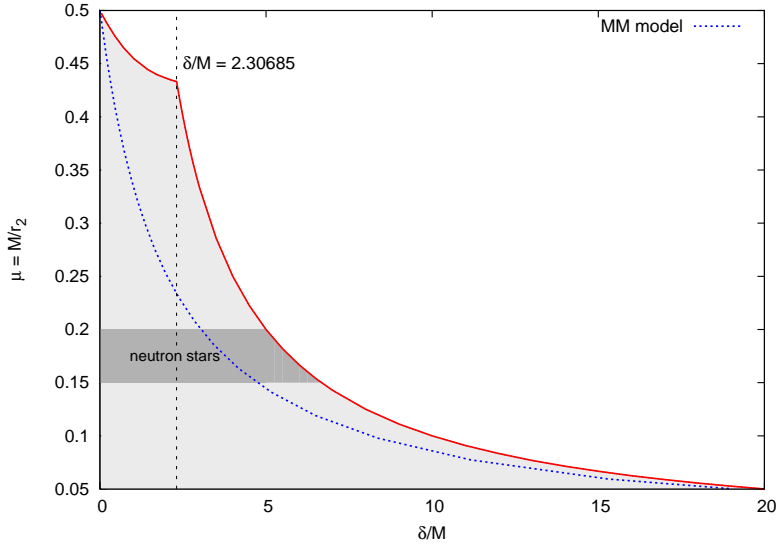}
  \caption{Limit on the compactness $\mu$ of the gravastar with the
    thickness of the shell $\delta$. This figure compares the results
    obtained for the MM model in the lower curve (see figure
    \ref{graf_novo}) and the results obtained for our model in the
    upper curve.}
  \label{eps}
\end{figure} 

Figure~\ref{eps} summarizes these results by showing the region of the
parameter space $(\delta, \mu)$ where equilibrium models can be
calculated. Equilibrium solutions for our model can be found in the
region below the solid line, and $\delta/M = 2.30685$ is the boundary
between the two distinct behaviors discussed above.  Indicated with a
dashed line is the corresponding threshold for the MM model
(\textit{cf.}  figure~\ref{graf_novo}). Overall, the data in the
figure reveals that our thick-shell model obeys similar but less
restrictive bounds than the MM model.

\section{Perturbation equations and numerical methods}

In order to assess the stability of gravastars against axial
perturbations we have followed the standard procedure to study
non-radial oscillations of stars~\cite{Chandra1,Chandra2,
Kokkotas,Nollert}. This approach differs from the one adopted in
ref.~\cite{DeBenedictis}, where the analysis of the axial
perturbations of a gravastar was made following the standard procedure
for Schwarzschild black holes (see~\cite{Nagar:2005ea} for a review)
in order to study only the asymptotic properties of the potential and
determine whether it has compact support and is everywhere positive.

We note that the pressure anisotropy in the shell does not play an
important role in the study of axial tensor perturbations since these
perturbations hardly excite fluid motions. As a result, the standard
analysis for stars with isotropic pressures can be used here. This is
no longer true for polar perturbations (for which the pressure
anisotropy plays an important role) and a study on radial
perturbations of stars with anisotropic pressures was recently
presented in~\cite{Gleiser1,Gleiser2}.

As for Schwarschild black holes or spherical relativistic stars, the
propagation of axial perturbations in the gravastar spacetime is
governed by the wave equation in a scattering potential
\begin{equation}
  \frac{\partial^2 \psi}{\partial r^{2}_*} - 
  \frac{\partial^2 \psi}{\partial t^2} = V_{\ell}(r)\psi\,,
  \label{wave1}
\end{equation}
where the ``tortoise'' coordinate is here
defined as
\begin{equation}
  r_* \equiv \int_0^r \rme^{({\lambda-\nu})/{2}}\rmd r\,,
  \label{r*}
\end{equation}
and the scattering potential is given by the expression
\begin{equation}
V_{\ell}(r) \equiv \frac{e^\nu}{r^3}\left[\ell(\ell+1)r+
	4\pi r^3(\rho-p_r)-6m\right]\,.
  \label{V(r)}
\end{equation}
The potential vanishes for $r \rightarrow \infty$ and diverges as
$1/r^2$ for $r \rightarrow 0$, as a result of the centrifugal term
proportional to $\ell(\ell+1)/r^2$. This is an important difference
that the potential $V_{\ell}(r)$ has with respect to its counterpart
for a Schwarzschild black hole, where $V_{\ell}(r) \rightarrow 0$ at
the horizon $r=2\,M$. Note also that $e^{\nu} \rightarrow 1$ for $r
\rightarrow \infty$ (the spacetime is the Schwarzschild one for
$r>r_2$) and, from eqs.~(\ref{g_00}) and (\ref{Gamma}), $e^{\nu}
\rightarrow (1-2M/r_2)e^{-\Gamma(r_2)}$ for $r \rightarrow 0$.

Unlike for black-hole spacetimes, both the tortoise coordinate and the
scattering potential are here not expressed through simple algebraic
relations but need to be computed as solutions of ordinary
differential equations. We have done so using a 4th order Runge-Kutta
method to obtain $\rme^\nu$ and $r_*$ from equations (\ref{g_00}) and
(\ref{r*}), respectively.

\begin{figure}[htp!]
  \includegraphics[angle=270,width=0.8\linewidth]{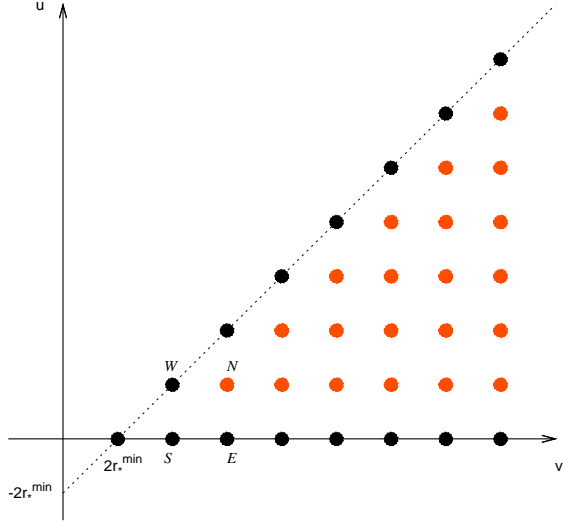}
  \caption{Diagram of the numerical grid and the domain of
    interest. The black points represent the grid points where the
    value of the field is known. The red points represent the grid
    points to be calculated.}
  \label{grid}
\end{figure}

Introducing now the ``light-cone'' (null) variables $u \equiv t-r_*$ and
$v \equiv t+r_*$, the wave equation (\ref{wave1}) can be written in the
more compact form as
\begin{equation}
  -4\frac{\partial^2\psi}{\partial u\partial v}(u,v) =
   V_{\ell}(r)\psi(u,v)\,,
  \label{wave2}
\end{equation}
and thus be integrated numerically. In particular, we have used a
variation of the method used in~\cite{Abdalla} and in many other works,
and derived a strict second-order accurate discretization given by
\begin{equation}
\psi_{_N} = (\psi_{_E}+\psi_{_W})\frac{16- \Delta^2
	V_{_S}}{16+\Delta^2 V_{_S}}- \psi_{_S} + {\cal
	O}(\Delta^4)\,,
\label{psi_n}
\end{equation}
where the indices $N$, $S$, $E$ and $W$ refer to the grid-points
defined as $N \equiv (u+\Delta,v+\Delta)$, $S \equiv (u,v)$, $E \equiv
(u,v+\Delta)$, and $W \equiv (u+\Delta,v)$.  Equation (\ref{wave2})
was then numerically integrated in the $(u,v)$ plane with the
algorithm (\ref{psi_n}) using a triangular grid limited by the lines
$r_* = r^{min}_*$ and $u = 0$, where $r^{min}_*$ is a small value
which we have set as $r^{min}_*=\Delta/2$. A schematic representation
of the null grid is presented in fig.~\ref{grid}, where the black
points represent the grid points where the solution is known, while
the red ones are those where the solution is to be calculated.

In a linear regime the eigenfrequencies of the gravastar are not
sensitive to the choice of the initial data and thus we have modelled
the initial perturbation with the simplest possible choice,
\textit{i.e.} a Gaussian pulse centred around $v_c$ and with witdth
$\sigma$
\begin{equation}
  \psi(u = u_0,v) = \exp\left[-\frac{(v-v_c)^2}{2\sigma^2}\right]\,,
\end{equation}
where, to satisfy the regularity condition at the origin, we set
\begin{equation}
  \psi(r^{min}_*,t) = \psi(u = v-v_0,v) = 0\,, \qquad~\forall t\,.
\end{equation}

During the integration of eq.~(\ref{wave2}) we extract the values of
the field $\psi$ along a line of constant $r_*$, \textit{i.e.}, at the
points $(u = v-2r_*,v)$. We let the field evolve for large values of
$t$, until the transient is over and possible contributions of the
higher overtones have died off. Only then, when the signal consists of
the fundamental mode (the slowest decaying mode) and of its overtones,
we can compute the real $\omega_{_R}$ and imaginary part $\omega_{_I}$
of its eigenfrequency with a least-squares fit of the function
\begin{equation}
\label{psi_fit}
\psi_{fit}(t) = 
	A_n{\rm exp}({\omega^n_{_{I}}t})\cos(\omega^n_{_R}t+\phi) \;,
\end{equation} 
where $A$ and $\phi$ are constant coefficients and $n$ represents the
order of the mode and this is not meant as an exponent. Once the
properties of the fundamental mode are known, the eigenfrequencies of
the overtones are obtained using the method proposed in~\cite{Davi},
which consists of subtracting from the numerical solution the fitted
function relative to the fundamental mode $n = 0$. This reveals the $n
= 1$ mode and the procedure can be iterated for obtaining higher
overtones, depending on their damping rate.  Of course, very rapidly
decaying modes cannot be obtained by this procedure, but this approach
is a very efficient and easy way to obtain the first overtones in a
signal, which are the ones expected to be detected from any
astrophysical sources.

The numerical setup described above has been first tested by
performing a perturbation analysis of Schwarzschild stars,
\textit{i.e.}, uniform-density spherical stars, obtaining results that
are in very good agreement with the values reported in the
literature~\cite{Chandra2,Star_modes}.

\section{Results}

Before discussing the results of the perturbative analysis, we note
that the axial potential given in (\ref{V(r)}) shows features which
are similar to those that are well-known in compact uniform-density
stars~\cite{Chandra2}. Quite generically, in fact, the potential
exhibits at least a minimum and a maximum, thus indicating the
possibility of having modes that are trapped in the potential well.

\begin{figure}[htp!]
  \includegraphics[angle=270,width=0.5\linewidth]{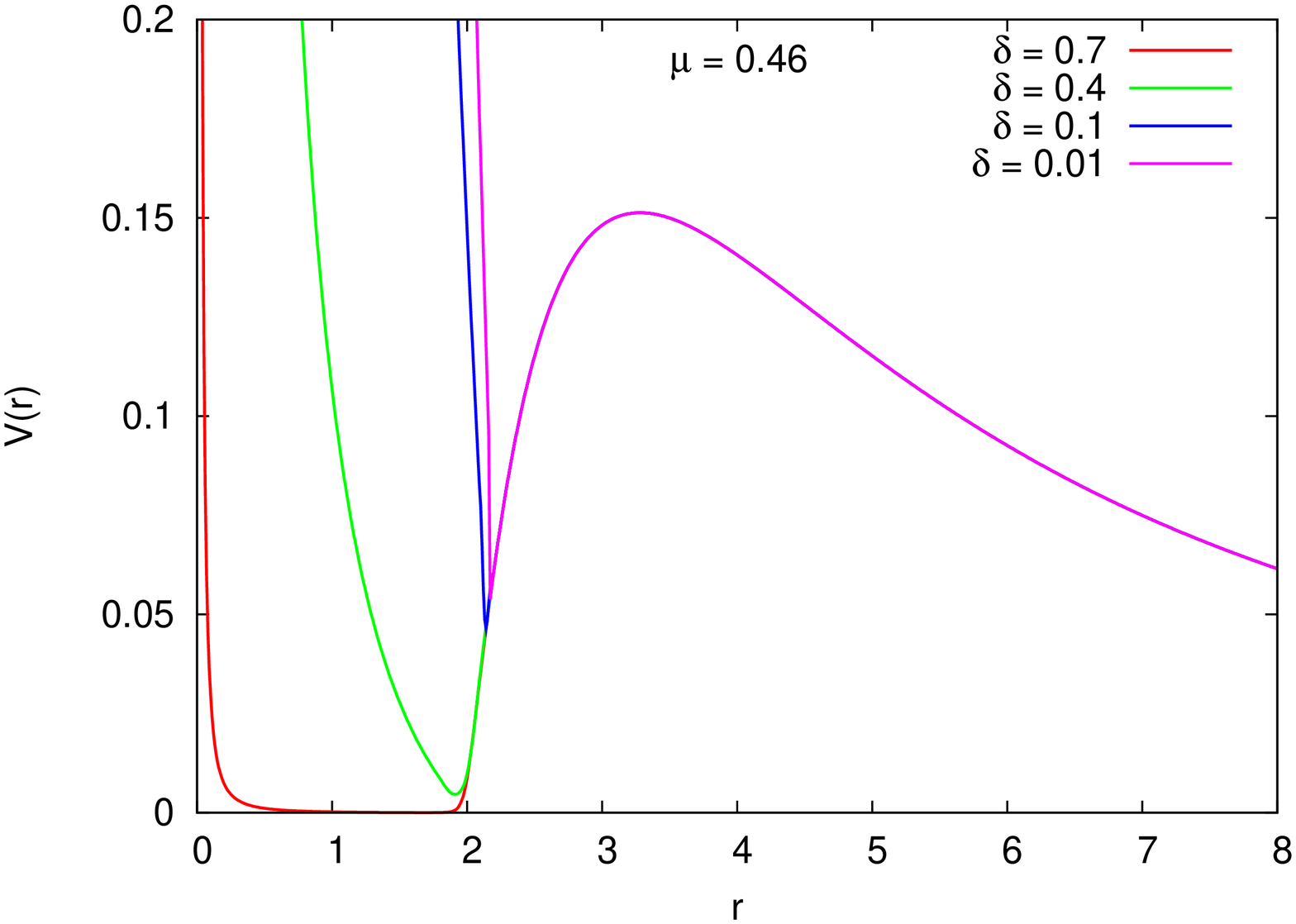}
  \hskip 0.2cm
  \includegraphics[angle=270,width=0.5\linewidth]{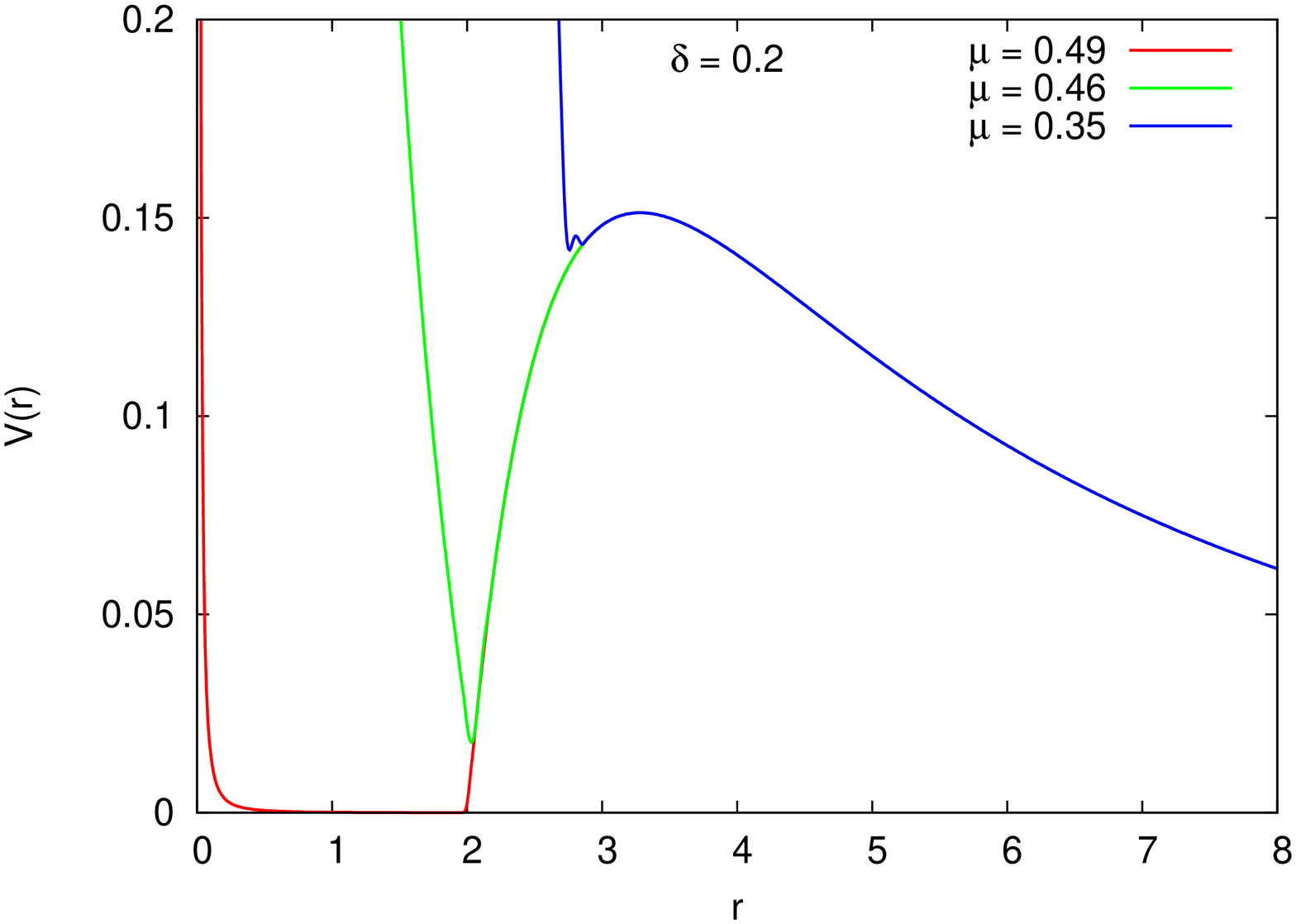}
  \caption{\textit{Left panel:} The scattering potential $V_{\ell}(r)$
    with $\ell = 2$ of gravastar models with the same compactness
    $\mu=0.46$ but increasing thickness of the shell. \textit{Right
    panel:} The same as in the left panel but gravastar models with
    the same shell-thickness $\delta = 0.2$ but increasing
    compactness.}
  \label{pot}
\end{figure}

In figure~\ref{pot} we present some typical examples of the scattering
potential $V_{\ell}(r)$ when $\ell = 2$ and which, in general, depends
on both the compactness $\mu$ (as for relativistic stars) and on the
thickness of the shell $\delta$. The left panel of figure \ref{pot},
in particular, shows the potential of gravastar models having the same
compactness $\mu=0.46$ but increasing thickness of the shell, while
the right panel reports how the potential changes for gravastar models
with the same shell-thickness $\delta = 0.2$ but increasing
compactness. Clearly, in the exterior region $V_{\ell}(r)$ is just the
Schwarzschild potential for axial perturbations and the depth of the
potential well in the interior region increases with both $\mu$ and
$\delta$. Furthermore, given the chosen EOS (\ref{eos}), it is also
possible to select the parameters $r_1$, $r_2$, and $M$ such that
$V_{\ell}(r)$ has two local minima very close to each other. This is
shown by the curve with $\mu = 0.35$ in the right panel of
figure~\ref{pot}.

\begin{figure}[htp!]
  \includegraphics[angle=270,width=0.5\linewidth]{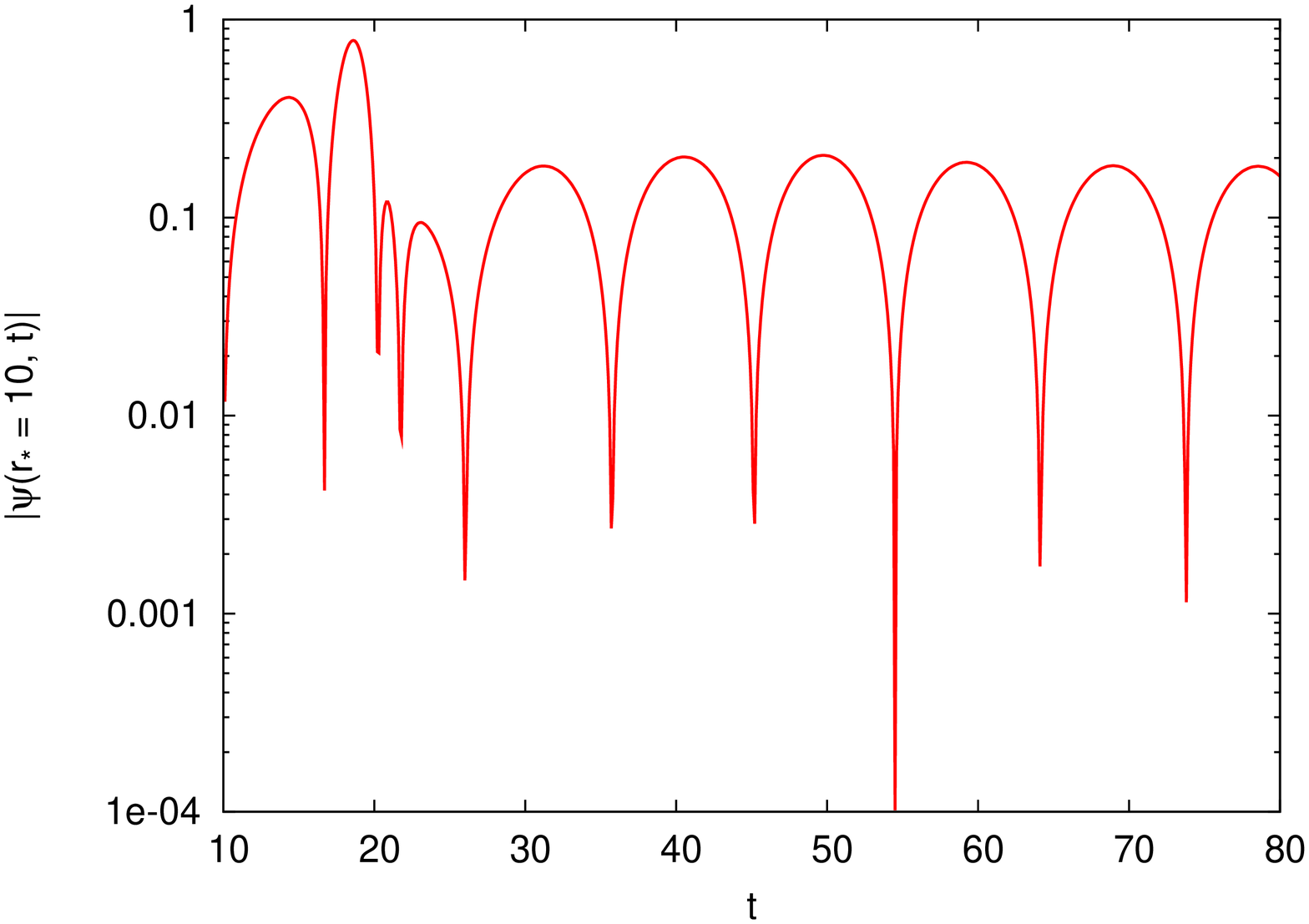}
  \hskip 0.2cm
  \includegraphics[angle=270,width=0.5\linewidth]{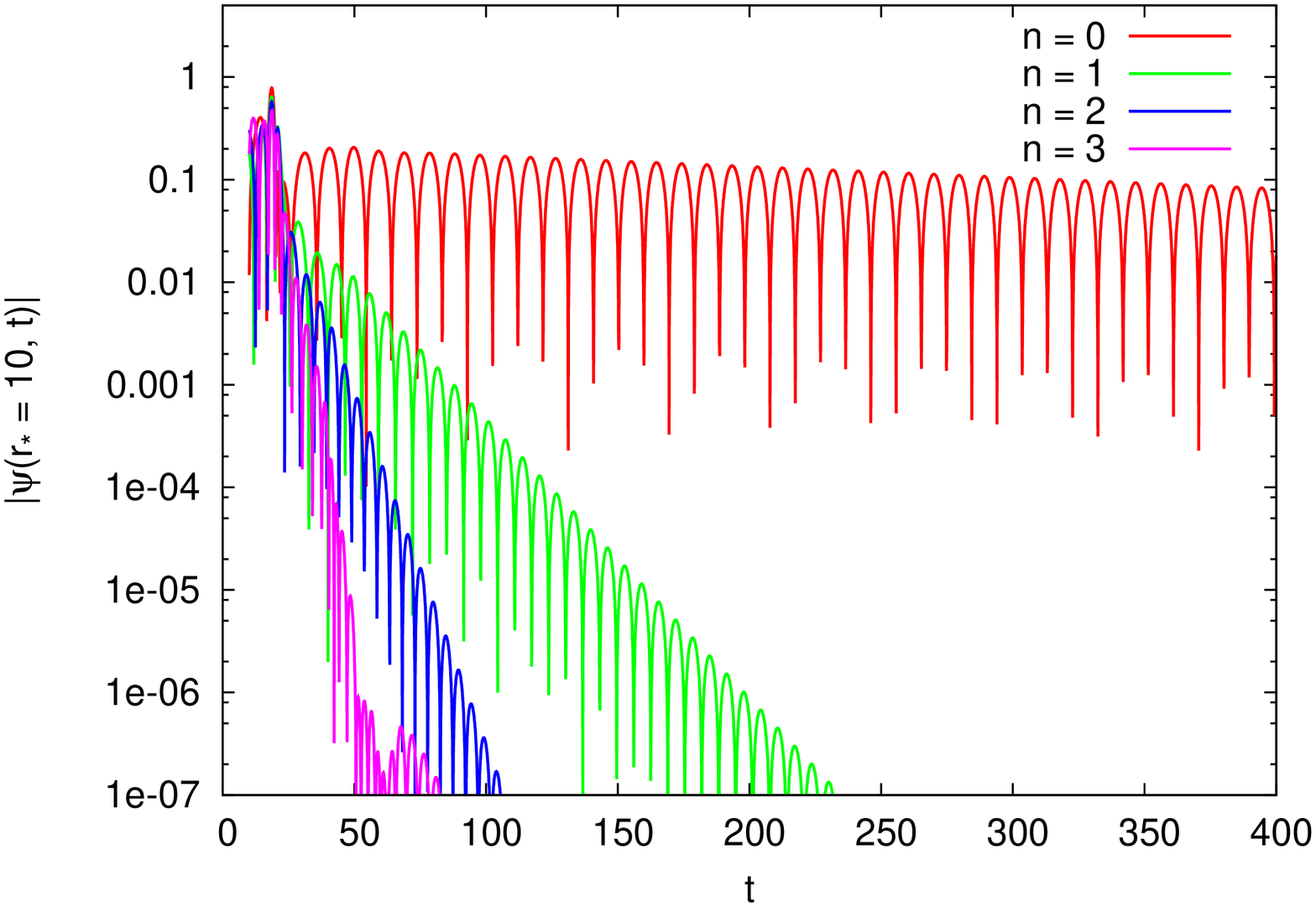}
  \caption{\textit{Left panel:} Evolution of the axial perturbations
    $\psi(t)$ as extracted at $r_*=10$ and during the initial stages
    of the scattering for a gravastar with $M=1$, $r_1=1.96$ and
    $r_2=2.26$. \textit{Right panel:} The same as in the left panel but
    when the contribution of the fundamental mode $n=0$, is
    distinguished from that of the overtones $n=1,2$ and 3.}
  \label{grav2}
\end{figure}

Following the procedure outlined in the previous Section, we have
integrated the perturbation equation (\ref{wave2}) for a large variety
of gravastar models differing in mass, compactness and thickness.
Figure~\ref{grav2} reports some typical results and shows the time
evolution of the perturbations for a representative gravastar with
$M=1$, and radii $r_1=1.96$, $r_2=2.26$. The left panel, in particular,
reports the total solution $\psi(t)$ as extracted at $r_*=10$ during
the initial stages of the scattering off the potential and allows one
to discern the ``beating'' caused by the superposition of overtones
and other rapidly decaying components of the signal.  Interestingly,
this beating of the fundamental mode with overtones becomes more
pronounced with increasing $\mu$ and $\delta$, underlining that the
contribution of the overtones increases (\textit{i.e.},~the modes
decay more slowly) with both $\mu$ and $\delta$ (\textit{i.e.},~ as
the the depth of the potential well increases). A similar relation
between the decay rate of the modes and the potential well is known
for the ``trapped modes'' of compact uniform-density stars
\cite{Star_modes}.

The right panel of figure~\ref{grav2}, on the other hand, shows the
evolution of the perturbations over a longer timescale and when the
overtones have been ``removed'' from the general signal using the
approach discussed in the previous Section. Note that, for increasing
$n$, the period of the oscillations becomes smaller, and the decaying
timescale shorter. Using a least-squares fit of the numerical data to
the \textit{ansatz}~(\ref{psi_fit}) we have computed the real and
imaginary parts of the QNM eigenfrequencies and the numerical results
for some typical cases are presented in table~\ref{table}, where they
are also compared to the corresponding values for a Schwarzschild
black hole of the same mass and a Schwarzschild star of the same
compactness. The tabulated values are all for an $\ell=2$ perturbation
and refer to the fundamental mode and to the first two overtones since
the third one does not usually have a sufficiently clear slope to
allow for an accurate measurement.
If necessary, an increase in spatial resolution would improve the
decay of higher overtones and allow for their computation.

\begin{table}
  \caption{\label{table}Some typical values obtained for the the first
    overtones of the gravastars QNM eigenfrequencies for axial
    perturbations with $\ell = 2$. The gravastars have $M=1$,
    $r_2=2.26$ and $\delta$ given in the Table. We also show for
    comparison the equivalent frequencies for a Schwarzschild black
    hole of the same mass and for a standard TOV star with uniform
    density with $M=1$ and $R=2.26M$ (note that for these stars one
    must have $R/M>2.25$~\cite{Chandra2}).}
  \begin{indented}
\bigskip
{
\item[]
      \begin{tabular}{lcccccc}
      \br
       & \centre{2}{$n=0$} & \centre{2}{$n=1$} & \centre{2}{$n=2$}\\
      &\crule{6}\\
      model
      & $\omega_{_{R}}$ & $-\omega_{_{I}}$ 
      & $\omega_{_{R}}$ & $-\omega_{_{I}}$
      & $\omega_{_{R}}$ & $-\omega_{_{I}}$ \\ 
      \mr
      $\delta=0.30$
      &$0.3281$&$2.481\mbox{e-}3$
      &$0.4865$&$6.264\mbox{e-}2$
      &$0.6534$&$1.590\mbox{e-}1$\\
      $\delta=0.35$
      &$0.2943$&$7.081\mbox{e-}4$
      &$0.4459$&$3.202\mbox{e-}2$
      &$0.5922$&$1.093\mbox{e-}1$\\
      $\delta=0.40$
      &$0.2575$&$1.543\mbox{e-}4$
      &$0.4011$&$1.227\mbox{e-}2$
      &$0.5384$&$5.814\mbox{e-}2$\\
      {\scriptsize Schwarzchild black hole}
      &$0.3737$&$8.896\mbox{e-}2$
      &$0.3467$&$2.739\mbox{e-}1$
      &$0.3011$&$4.783\mbox{e-}1$\\
      {\scriptsize Schwarzchild star}
      &$0.1090$&$1.239\mbox{e-}9$
      &$0.1484$&$3.950\mbox{e-}8$
      &$0.1876$&$5.470\mbox{e-}7$\\
      \br
    \end{tabular}
}
  \end{indented}
\end{table}

\begin{figure}[htp!]
  \includegraphics[angle=270,width=0.5\linewidth]{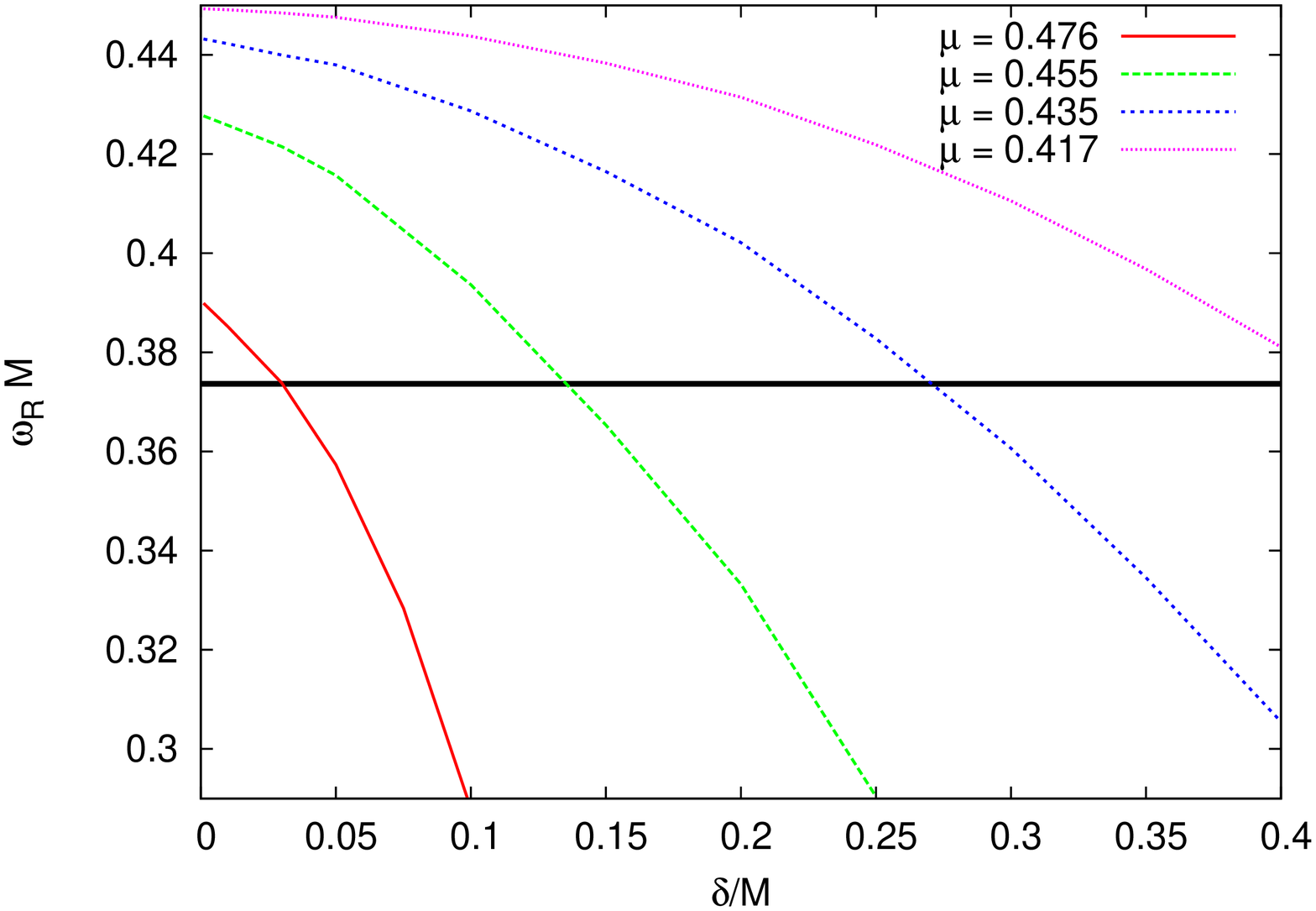}
  \hskip 0.2cm
  \includegraphics[angle=270,width=0.5\linewidth]{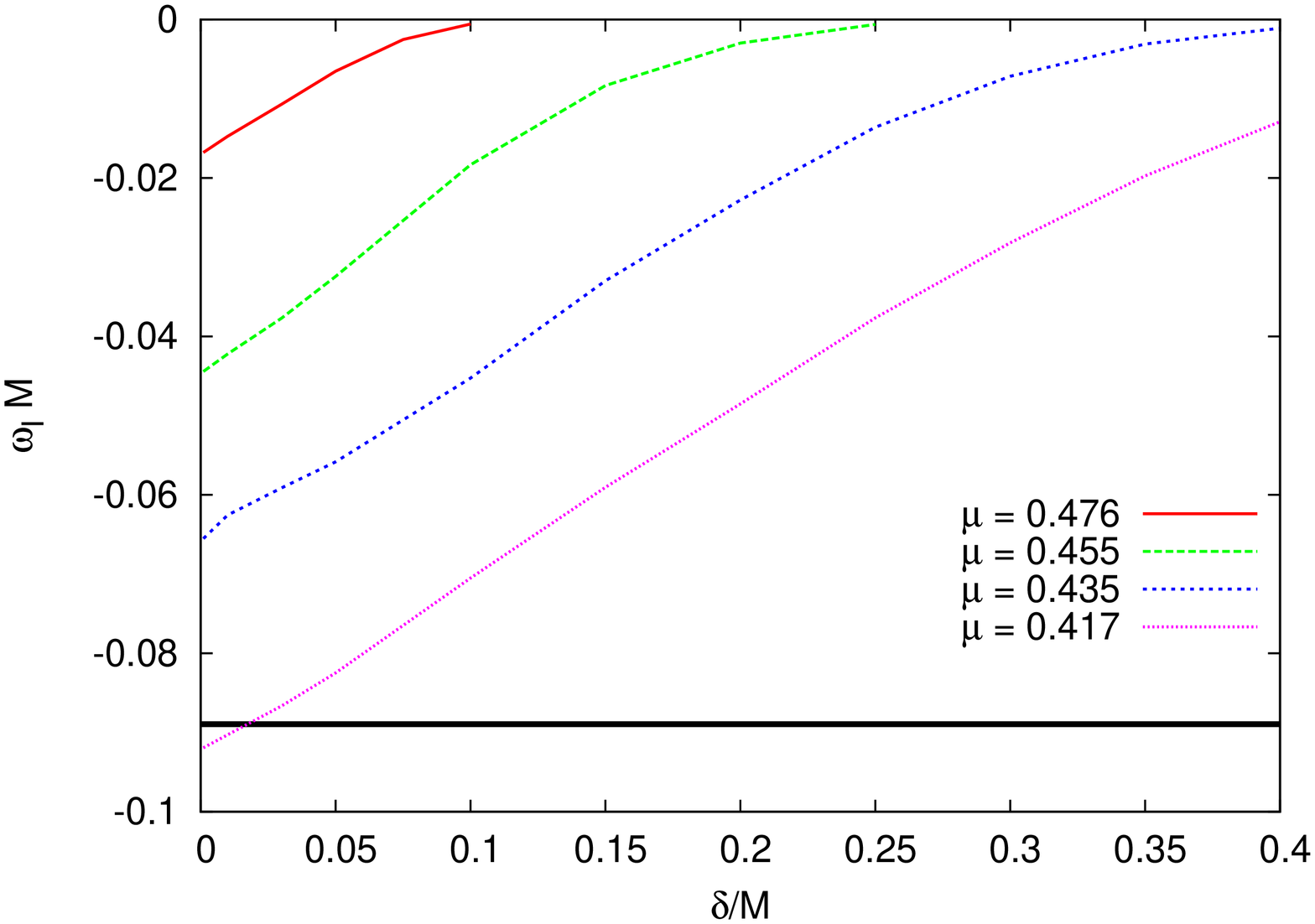}
  \caption{\textit{Left panel:} Behaviour of the real part
    $\omega_{_{R}}$ of the $\ell=2$, $n=0$ QNM eigenfrequencies as
    functions of $\delta$ for different values of $\mu$. \textit{Right
    panel:} The same as in the left panel but for the imaginary part
    $\omega_{_{I}}$ of the eigenfrequencies. In both panels, the thick
    horizontal lines represent the corresponding frequencies for a
    Schwarzschild black hole.}
  \label{qnm}
\end{figure}

Figure~\ref{qnm} is the most important of this paper and it reports
the variation of the $\ell=2$, $n=0$ QNM eigenfrequencies as a
function of the parameters of the gravastar: $\mu$ and $\delta$. The
two panels report separately the behaviour of the real and imaginary
parts and indicate that both the period of the oscillations
$\tau_{_{R}} \equiv 2\pi/\omega_{_{R}}$ and the the damping time
$\tau_{_{I}} \equiv 2\pi/|\omega_{_{I}}|$ increase with increasing
$\delta$ and $\mu$. The two thick horizontal lines represent instead
the corresponding frequencies for a Schwarzschild black hole of the
same mass, \textit{i.e.}, $M\omega_{_R} = 0.37367$ and $M\omega_{_I} =
-0.08896$~\cite{Kokkotas}. Clearly, while it is always possible to
select the thickness and compactness of the gravastar such that it
will have the same oscillation frequency of a black hole with the same
mass (\textit{cf.}, left panel), the corresponding decaying time will
be different and about one order of magnitude larger (\textit{cf.}, right
panel). Stated differently, a gravastar and a black hole with the same
mass cannot have the same complex QNM eigenfrequency when subject to
axial perturbations. This result, which we have here presented for the
fundamental mode, can be shown to be true also for the first two
overtones.

\section{Conclusions}

Although the gravastar model of Mazur and Mottola~\cite{Mazur}
represents an ingenious solution of the Einstein equations in
spherical symmetry, it has also challenged one of the most cherished
foundations of modern astrophysics: \textit{i.e.,} the existence of
astrophysical black holes. Gravastars, in fact, can be constructed to
be arbitrarily compact, with an external surface which is only
infinitesimally larger than the horizon of a black hole with the same
mass. As a result, the electromagnetic emission from the surface of a
gravastar will suffer of essentially the same gravitational redshift
as that of a black hole, thus making it difficult, if possible at all,
to distinguish the two when only electromagnetic radiation is
available.

Without entering the relevant debate about the physical processes that
would lead to the formation of a gravastar or the astronomical
evidence in support of their existence~\cite{Broderick}, we have here
considered two more fundamental questions: Is a gravastar
\textit{stable} against generic perturbations? If so, can an external
observer \textit{distinguish} it from a black hole?  The short answers
to these questions are that: a gravastar \textit{is stable} to axial
perturbations and indeed \textit{it is possible} to distinguish it
from a black hole if gravitational radiation is produced.

To reach the first of these conclusions we have constructed a general
class of gravastar models that extends the one proposed by Mazur and
Mottola by replacing the infinitesimal shell of matter with one having
finite size $\delta$ and variable compactness $\mu$. These equilibrium
solutions of the Einstein equations have then been analyzed when
subject to axial perturbations and the eigenfrequencies of the
corresponding QNMs have been computed explicitely. For all of the
cases considered, the imaginary part of the eigenfrequencies has
always been found to be negative, thus indicating the stability of
these objects with respect to this type of perturbations.

To reach the second conclusion, instead, we have shown that the QNM
spectra of a gravastar and that of a black hole of the same mass
differ considerably. In particular, while it is always possible to
select $\delta$ and $\mu$ such that the gravastar has the same
oscillation frequency as that of a black hole with the same mass, the
corresponding decaying time will be different. As a result, the
gravitational radiation produced by an oscillating gravastar can be
used to distinguish it, beyond dispute, from a black hole of the same
mass.

We plan to extend our stability analysis also to polar perturbations
and determine whether or not these intriguing objects possess modes of
oscillation that do not have a counterpart in compact relativistic
stars and may therefore hint to new solutions of the Einstein
equations.

\ack 
LR dedicates this work to Richard A. Matzner in occasion of his
65$^{\rm th}$ birthday.  Support comes from the Funda\c c\~ao de
Amparo \`a Pesquisa no Estado de S\~ao Paulo (FAPESP), the Deutsche
Akademische Austauschdienst (DAAD) and the Max-Planck-Gesellschaft
(MPG).

\end{document}